\documentclass[graybox,natbib,nosecnum]{svmult}
\bibpunct{(}{)}{;}{a}{}{,} % suppress commas between author-names and year

\pdfoutput=1   %forces use of pdflatex. Disable if you prefer to use .eps or .ps figures.
\usepackage{mathptmx}       % selects Times Roman as basic font
\usepackage{helvet}         % selects Helvetica as sans-serif font
\usepackage{courier}        % selects Courier as typewriter font
\usepackage{type1cm}        % activate if the above 3 fonts are
\usepackage{amssymb}         % allows index generation
\usepackage{makeidx}         % allows index generation
\usepackage{graphicx}        % standard LaTeX graphics tool
\usepackage{multicol}        % used for the two-column index
\usepackage[bottom]{footmisc}% places footnotes at page bottom
\usepackage[normalem]{ulem}	% for strike-through of text with \sout{}  
\usepackage{hyperref}  %for hyperlinks

\usepackage{soul}   % for high-lighting of text
\makeindex             % used for the subject index
\begin{document}

\title*{Chemical signatures of planet engulfment events in Sun-like stars}
\author{Spina Lorenzo}
\institute{INAF - Osservatorio Astrofisico di Arcetri, Largo E. Fermi 5, 50125, Firenze, Italy, \email{lorenzo.spina@inaf.it}
%\and First (Middle) Family name of second author \at Name, Address of Institute \email{name@email.address}
}
\maketitle

\abstract{The observational evidence that planetary systems can be very different from each other, suggests that their dynamical histories were very diverse, probably as a result of a strong sensitivity to the initial conditions. Severe dynamical processes can drive the orbital decay of planets or planetesimals ending in their accretion onto the host star. When this material enters the star, it is rapidly dissolved in the stellar envelope, altering the star's chemical pattern in a way that mirrors the composition observed in rocky objects. Indeed, chemical signatures of planet ingestion has been found in an increasing number of Sun-like stars. These observations carry substantial implications for the field of exoplanet science, as they are entirely detached from both specific biases associated with exoplanet detection techniques and assumptions made in n-body numerical simulations of systems' evolution. For instance, signatures of planet engulfment events suggest that a non-negligible portion of planetary systems has undergone highly dynamic histories, ultimately resulting in the fall of planetary material into the host star. Also, these studies open to the possibility of using chemical abundances of stars to identify which ones are the most likely to host analogues of the calm Solar System.
}

\section{Introduction}
% As a guideline, your contribution should be 5000-7000 words (or as agreed with the section editors), including references.
% 721 words
% 1507 words
% 773 words
% 

Although the detectability of Jupiters orbiting Sun-like stars was anticipated already 70 years ago \citep{Struve52}, the discovery of 51 Peg b \citep{Mayor95} came as a significant surprise because it implied something that was totally unexpected. In fact, Jupiter-mass planets were supposed to form at large radii, 5–10 AU where most of the disk mass resides. Thus, the extremely short orbit of a very massive planet implied that significant orbital migration can occur around Sun-like stars, unlike to what we have deduced from our own Solar System.

After almost 30 years from the groundbreaking discovery of 51 Peg b, we have made significant progress in the detection of exoplanets orbiting stars similar to our Sun. Prodigious efforts in this field have yielded the realisation that nearly every star in the universe is likely to host at least one planet. Furthermore, it has been found that roughly a half of Sun-like stars possess rocky planets residing within their habitable zones. However, one of the most undeniable and striking evidence that has emerged from exoplanet exploration is the existing great diversity among planetary systems, suggesting that dynamical processes play a key role within these systems since the initial formation of planetesimals \citep[e.g.][]{Izidoro17}. The fact that some systems have undergone complex phases of dynamical evolution is attested to by the presence of various architectures which are strikingly different from that of the Solar System: systems with close-in gaseous planets or super-earths \citep[e.g.][]{Howard12}, systems with planets on highly eccentric orbits \citep[e.g.][]{Kane12}, misaligned or even counter rotating with respect to the spin axes of their hosting star \citep[e.g.][]{Naoz11}, dusty debris disks formed through exoplanet collisions \citep[e.g.][]{Kenyon16}, or even interstellar exoplanets \citep[e.g.][]{Mroz18}.

Dynamical processes in the most chaotic systems have the possibility to destabilise planets from their orbits, forcing them to plunge into the host star. These planet engulfment events around Sun-like stars are predicted by numerical models of systems' evolution \citep[e.g.][]{Mustill15} and they can also occur during the main-sequence phase \citep[e.g.][]{Church20,Rodet23}. The plausibility of this scenario is now supported by the first direct observation of a Sun-like star swallowing one of its own planets \citep{De23}.

Planet engulfment events involve the chemical assimilation of the planetary material into a star’s external layer \citep{Pinsonneault01}. As a result, the metallicity of the parent star can be enhanced in a way that mirrors the composition of the engulfed rocky object, with rocky-forming elements - such as iron - resulting more abundant than what they would be otherwise \citep[][]{Laughlin97}. 

%Therefore, if a star identical to the Sun has a convective envelope subtending 2$\%$ of its mass, even a few Earth masses would produce a chemical enhancement that is both significant and observable \citep{Chambers10}. Indeed, spectroscopic studies of Sun-like stars have measured anomalies in the chemical patterns of their atmospheres that are interpreted as the result of planet engulfment events \citep[e.g.][]{Spina21}.

All these considerations lead us to some intriguing questions. How often do Sun-like stars swallow their own planets? What is the occurrence of unstable planetary systems? How rare are ordered and calm planetary systems like our own Solar System? Can we detect signatures of planet ingestion in the chemical composition of Sun-like stars? If so, can we use the chemical pattern of a star to infer architecture of its planetary system? 

This short review outlines how all these questions are being tackled with theoretical and observational studies of stellar atmospheres, including that of our Sun. %The manuscript is structured as it follows. First, Section~\ref{Sec:theoretical}...

%It is likely that in systems with evidence of a dynamical past, part of the planetary material has fallen into the hosting star (e.g., Martinez et al. 2019; Liu et al. 2018), polluting its atmosphere and producing a significant increase in the stellar metallicity, which can be reliably detected (e.g., Spina et al. 2015, 2018; Oh et al. 2018; Tucci Maia et al. 2019). 

%Studies on binary systems (e.g., Tucci Maia et al. 2014; Biazzo et al. 2015; Ramírez et al. 2015; Teske et al. 2015, 2016; Saffe et al. 2016, 2017; Oh et al. 2018) and open clusters (Spina et al. 2015) have demonstrated that differences can exist in the composition of stars belonging to the same association and that these chemical anomalies are typically larger for the most refractory elements.

\section{Rocky bodies polluting atmospheres of Sun-like stars}
\label{Sec:theoretical}

This Section explains the ways and conditions in which planet engulfment events can produce detectable chemical signatures in the atmospheres of stars. Throughout the entire manuscript the usual astronomical scale for logarithmic abundances is adopted. In this notation hydrogen abundance is defined to be A(H)~=~12.00, while the abundance of any other X-element is  A(X)~=~log(N$_{\rm X}$/N$_{\rm H}$)~+~12, where N$_{\rm X}$ and N$_{\rm H}$ are the number densities of X-element and hydrogen, respectively. Furthermore, the abundance ratios relatively to the Sun is defined as [X/H]~=~A(X)~-~A$_{\rm \odot}$(X) and [X/Fe]~=~[X/H]~-~[Fe/H]. 

As a toy model, it is assumed that the full amount of planetary material fallen onto the star is entirely diluted within the external convective zone and there it stays forever polluting the stellar atmosphere with specific elements. In this simple model, the abundance variation of a given X-element in the stellar convective zone is $\Delta$[X/H]~=~[X/H]$^f$~-~[X/H]$^i$~=~log~N$_X^f$/N$_X^i$, where [X/H]$^i$ and [X/H]$^f$ are the abundances before and after the planet engulfment event. Similarly, N$_X^i$ and N$_X^f$ are the number densities of the X-element before and after the event. More specifically, N$_X^i$ is equal to the number density within the pristine stellar convective zone while N$_X^f$ can be assumed to be equal to N$_X^i$ plus that of the accreted mass. Therefore, the abundance variation of the X-element can be expressed as

\begin{equation}
\label{Ax}
    \Delta \left[\frac{X}{H}\right] = log_{10} \frac{M_X^{cz} + M_X^{pl}}{M_X^{cz}},
\end{equation}

where M$_{\rm X}^{\rm cz}$ is the mass associated to the X-element enclosed within the original convective zone and M$_{\rm X}^{\rm pl}$ is the additional mass brought by the planet. Models of chemical pollution based on these simple calculations have been developed by, e.g., \citet{Chambers10, Mack14, Spina15, Galarza21}.

%\subsection{Planetary material accreted onto the star (M$_{\rm X}^{\rm pl}$)}
%\label{Sec:material}

The mass of the X-element contaminating the stellar atmosphere is expressed as 

\begin{equation}
    M_X^{pl} = M_{pl} \times \frac{a_x \times 10^{A_{pl}(X)}}{\sum_x a_x \times 10^{A_{pl}(X)}},
\end{equation}

where M$_{\rm pl}$ is the total mass of the planet, A$_{\rm pl}$(X) is the abundance of the X-element within the planet, and a$_X$ is the atomic weight of the X-element. The chemical abundances A$_{\rm pl}$(X) are typically assumed to be those of chondritic compositions in the Solar System \citep{Wasson88,Lodders09}, that of the Earth \citep{McDonough03}, or a combination between them. 

\begin{figure*}
    \centering
    \includegraphics[width=1.\textwidth]{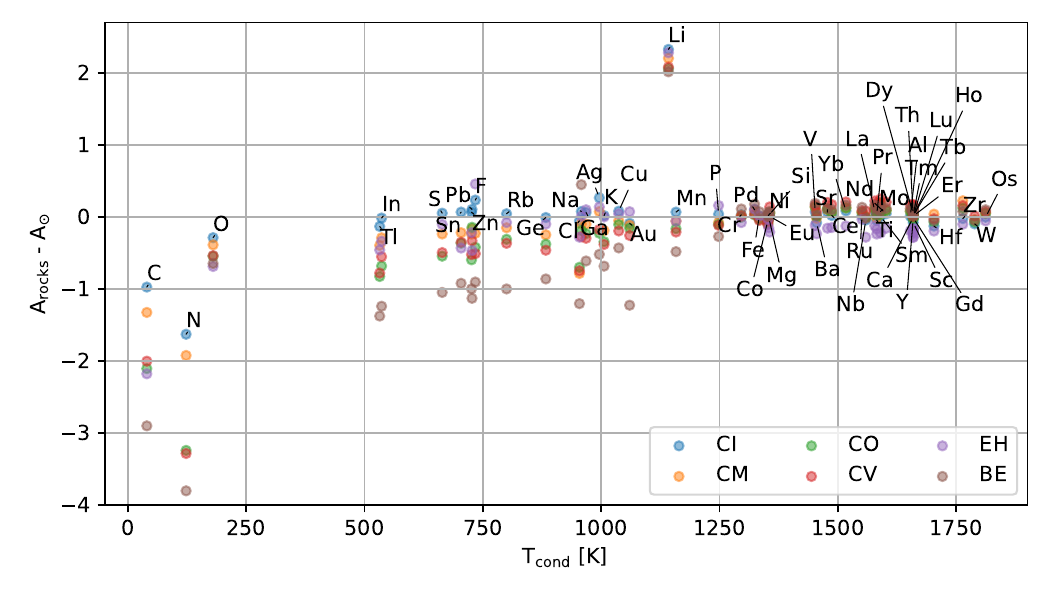}
    \caption{Chemical composition of chondritic meteorites \citep[carbonaceous chondrites: CI, CM, CO, and CV;  enstatite chondrites: EH;][]{Wasson88} and the bulk of the Earth \citep[BE][]{McDonough03} relatively to the chemical composition of the Sun \citep{Asplund21}. Differential abundances A$_{\rm rocks}$-A$_{\odot}$ are plotted against the elemental condensation temperature T$_{\rm cond}$.}
    \label{fig:rocky_composition}
\end{figure*}

The chemical pattern of the planetary material dictates the distinct features of the chemical imprint that a planet engulfment event can leave on the stellar atmosphere. In this regard it is useful to introduce the notion of condensation temperature T$_{\rm cond}$ which is widely used in planetary science to classify elements based on their ability to solidify into dust and rocks. Species condensing at very high temperatures  (i.e., T$_{\rm cond}$$>$1000~K) - such as Fe, Ti, Al, Sc - are called  {\it refractory elements}. Within protoplanetary disks, they are the first elements that solidify to form the bulks of planetesimals, planets, meteorites and all the rocks populating planetary systems. Conversely, the so-called  {\it volatile elements} - such as C, O - need extremely low temperatures to solidify  (i.e., T$_{\rm cond}$$<$1000~K) thus they prefer to remain in the gaseous phase. As a result, {\it volatile elements} do not significantly contribute to the formation of rocks, even though they can be retained within the atmospheres of planets. When we compare the chemical pattern of rocks A$_{\rm rocks}$(X) to the composition of the Solar atmosphere A$_{\odot}$(X), we observe that the differential abundances A$_{\rm rocks}$(X)-A$_{\odot}$(X) somehow correlate with T$_{\rm cond}$, as it is shown in  Fig.~\ref{fig:rocky_composition}. While refractory elements are present with similar abundances within rocks and in the Sun, as we move towards elements with lower T$_{\rm cond}$ (i.e., volatiles) we find that they are ever less abundant in rocks. A planet engulfment event is expected to leave on the stellar atmosphere a chemical signature that mirrors the chemical pattern shown in Fig.~\ref{fig:rocky_composition}, with a significant enhancement of the refractory elements, while no (or very little) change is expected for the volatiles.

Figure~\ref{fig:rocky_composition} also shows that there is one remarkable outlier to the smooth relation that links A$_{\rm rocks}$-A$_{\odot}$ to T$_{\rm cond}$: lithium. All stars are born with a similar amount of Li in their atmospheres. However, Sun-like stars burn most of their initial endowment during theirs first few 100 Myr. Afterwards, the Li depletion rate significantly slows down, even though it continues at a very small regime during the entire main-sequence phase. In contrast, rocks in planetary systems fully retain their initial Li reservoir across their entire lifetime. That explains why chondrites and the Earth are 2 dex richer in Li than the Sun (see Fig.~\ref{fig:rocky_composition}), in fact the Sun has already depleted $\sim$2 dex of its initial Li abundance. Therefore, if the engulfment event occurs when the star is older than a few 100 Myr, the rocky body is expected to refurbish the depleted Li reservoirs in the stellar atmosphere, hence a significant increase of the Li abundance is expected \citep[][]{Sandquist02}. Instead, if the rocky material is accreted by a much younger star, the Li carried by the planet will be burnt together with the stellar Li. In this case, after enough time from the engulfment event, the Li abundance of the polluted star would not be any different from that expected by a normal star. In conclusion, the rise of Li abundance due to a planet engulfment event strongly depends on the age of the star at the time of the event and also on how much Li the star has depleted since then. The recent work by \citet{Sevilla22} shows results from computations of the level of Li enrichment due to planet engulfment events.

%Among all the elements shown in Fig. X, Lithium stands out for its peculiar behaviour. Typically refractory elements have abundances that are similar to those of the Sun. That is expected, as these abundances reflect the chemical pattern of the proto-Solar material from which the Sun and the planets were formed. Instead, the abundance of Li in rocks is much higher than that of the Solar atmosphere. The origin of this discrepancy must be traced back to the fact that Sun-like stars deplete the Lithium in their atmospheres as times goes on. The Lithium depletion rate depends on the extension of the stellar convective zone at each stage of its evolution, thus it is also strongly conditioned by the stellar mass. However, the depletion of Lithium is very well studied in stars similar to our Sun. Fig. XXX shows that a star that is 4.5 Gyr old has already lowered its Li abundance of 1-1.5 dex. Given the fact that Li depletion only occurs in stars but not in planets, the rocky bodies in our Solar System are significantly richer in Li than the Sun.

%\subsection{The stellar convective zone (M$_{\rm X}^{\rm cz}$)}

Besides the chemical composition of the mass accreted onto the star, the other key ingredient of Eq.~\ref{Ax} is the mass of the X-element enclosed within the stellar convective zone (M$_{\rm X}^{\rm cz}$). This latter it is a function of both the chemical abundance A$_{\rm X}$ measured in the stellar atmosphere and the extent of the convective zone at the time $\tau$ of the planet engulfment event, i.e., M$_{\rm cz}$$($$\tau$):

\begin{equation}
    M_X^{cz} = M_{cz}(\tau) \times \frac{a_x \times 10^{A_X}}{\sum_x a_x \times 10^{A_X}}.
\end{equation}

%where a$_X$ is the atomic weight of the X-element. 

%Similarly M$_X^{pl}$ is function of the total planetary mass M$_{pl}$ and its individual abundances A$_X^{pl}$. Namely, 

%\begin{equation}
%    M_X^{pl} = M_{pl} \times \frac{a_x \times 10^{A_X^{pl}}}{\sum_x a_x \times 10^{A_X^{pl}}}.
%\end{equation}

%From these we derive the new mass of the X-element enclosed enclosed in the polluted atmosphere after the planet engulfment event:

%\begin{equation}
%    M_X^{cz+pl} = M_X^{cz} + M_X^{pl},
%\end{equation}

%and therefore the abundance variation:

%\begin{equation}
%\label{delta}
%    \Delta A_X = log_{10} \left(\frac{M_X^{cz+pl}}{M_{CZ}(\tau)} \times \frac{\sum_x a_x \times 10^{A_X}}{a_x}\right) - A_X.
%\end{equation}

It is well known that, during the early phase of contraction, solar-type stars undergo a process of internal readjustment in which the extended CZ retreats toward the surface until it stabilises as soon as the star reaches the zero-age main-sequence \citep[e.g.][]{Palla93}. Sun-like stars pass from being fully convective (i.e., M$_{\rm cz}\sim$1M$_{\odot}$) to having an extremely thin external convective layer which is just a few percent of the stellar mass. Therefore, besides the amount of polluting material, also M$_{\rm cz}$$($$\tau$) plays a fundamental role in determining the level of chemical enrichment. More specifically, when M$_{\rm cz}$ $\gg$ M$_{\rm pl}$ the stellar convective zones can easily dilute the full amount of rocks acquired from the planetary system without changing the atmospheric chemical composition. Otherwise, a much thinner convective zone can get significantly polluted even by a modest amount of rocky material. In that case a chemical signature of the engulfment event could be observed. 

\begin{figure*}
    \centering
    \includegraphics[width=1.\textwidth]{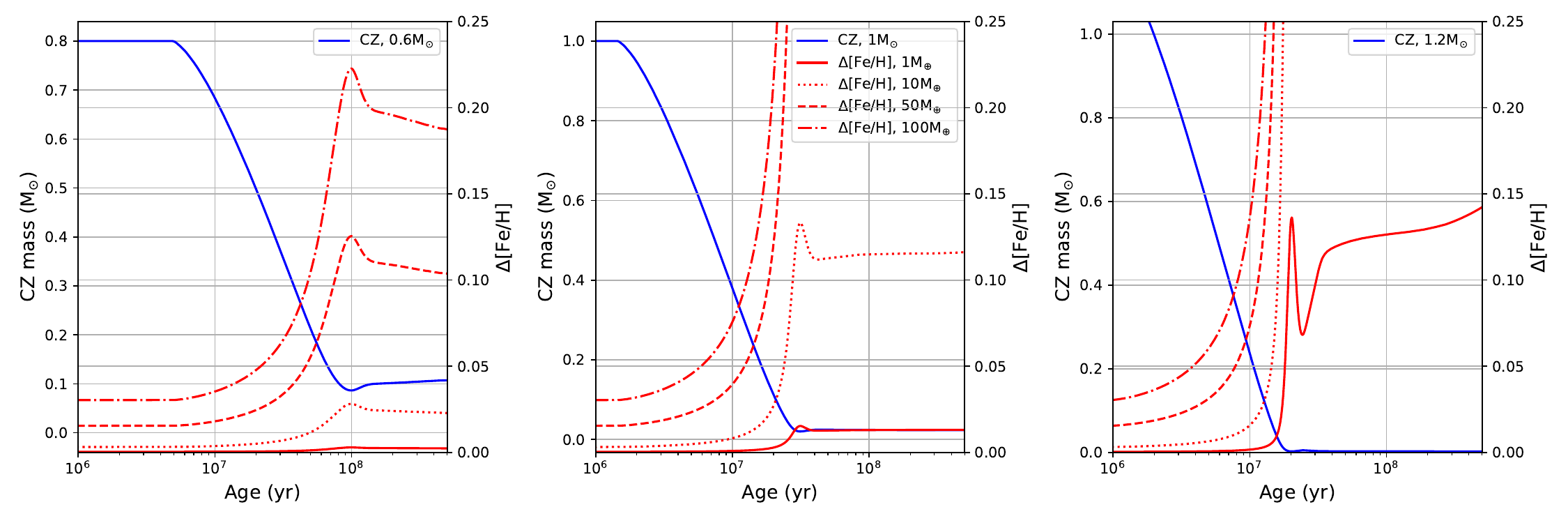}
    \caption{Variation of the mass enclosed within the convective zone for a 0.6, 1 and 1.2 M$_{\odot}$ star as a function of time (blue solid line). The three panels also show the variation in iron abundance $\Delta$[Fe/H] produced by the ingestion of 1, 10, 50, and 100 M$_{\oplus}$ of rocky material occurred at different sizes of the convective zone (red lines). YaPSI stellar models are used \citep{Spada17}. The rocky material is assumed to have the same composition of Earth's core \citep[][]{McDonough03}.}
    \label{fig:enrichment}
\end{figure*}

Both the typical shrinking timescales of the CZ and its mass at the main-sequence are function of the full stellar mass: more massive stars undergo a faster shrinking of their CZs and the mass enclosed in the external layer during the main-sequence phase is smaller. In Fig.~\ref{fig:enrichment} it is shown the variation in CZ mass (solid blue line) during the first 400 Myr of the evolution for a 0.6, 1.0, and 1.2 M$_{\odot}$ star. The figure also illustrates the time variation of the atmospheric iron content of the star (red lines) resulting from the accretion of objects with different masses during the pre-main-sequence contraction phase: 1, 10, 50, and 100 M$_{\oplus}$ of rocky material. As we can see, a fixed mass of rocks can produce extremely different iron enhancements depending on the stellar age owing to the rapidly evolving decrease of the CZ mass.

\subsection{The role of mixing processes}
\label{Sec:mixing_proc}

The enhancement of refractory elements within the engulfing star's convective region may be weakened over time by internal mixing processes and - in particular - by thermohaline convection \citep[e.g.][]{Ulrich72,Vauclair04,Theado12,Garaud21}. The freshly accreted metals diluted in the small convective zone and floating above the radiative layer of pristine stellar material create a mean-molecular-weight gradient which is highly unstable against convection. This instability induces blobs of metal-enriched matter begin to fall down from the convective zone and exchange heavy elements with the metal-poorer material of the stellar interior. Thus, this process is supposed to remove the polluting material from the stellar atmosphere. However, numerical simulations and semi-analytical estimation have shown that thermohaline convection can hardly completely erase the excess of metals on the surface of Sun-like stars. The recent simulations by \citet{Behmard23} have shown that $\gtrsim$1~M$_{\odot}$ stars can retain a significant excess of metals on their surfaces for at least one or more Gyr after the engulfment event (see their Fig. 1). The larger the stellar mass and the longest the metals can survive on the external layer. Also, since more massive stars have thinner external layers that get easily polluted, the larger the stellar mass and the larger the chemical signature is left by a planet engulfment event. More specifically, a 1~M$_{\odot}$ star that has engulfed a 10~M$_{\oplus}$ bulk-Earth composition planet is expected to retain for at least 1 Gyr a $\Delta$[Fe/H]$\sim$0.05~dex. This abundance variation is certainly observable through $high-precision$ methods of analysis such as the differential line-by-line analysis \citep{Nissen18}, which allows Fe abundance determinations at sub-0.01 dex precision for Sun-like stars \citep[][]{Ramirez14,Nissen15,Spina16a,Spina16b,Bedell18,Spina18,Casali20}. Instead, a 1.2~M$_{\odot}$ star will retain a $\Delta$[Fe/H]$\sim$0.15~dex for most of its main-sequence phase. Notice however that these values result from a planet engulfment event occurring at the stellar zero-age-main-sequence (ZAMS). The same authors show that later engulfment events would result in even longer-lasting and less-diluted chemical signatures. 

In conclusion, chemical signatures of planet engulfment events are more likely to be observed in stars whose masses are $\gtrsim$1~M$_{\odot}$. That is not only because these have thinner convective envelopes that can get more easily polluted, but also because these signatures are longer-lived and not completely erased by thermohaline convection. 

However, the study of thermohaline mixing in stars is far from trivial \citep[][]{Lattanzio15} as the exact amount of metals that may remain in the stellar outer layer depends on key parameters, which cannot be precisely constrained - such as inertia, size, and length of the metallic blobs or the elemental diffusivity in the stellar interior - and on numerical assumptions like spatial and time resolution adopted in the computations. On top of that, other physical agents (e.g., rotation, magnetic fields, shears) that are present in stars and that are typically neglected in these studies may strongly affect the efficiency of mixing processes \citep[][]{Garaud21}. As a consequence, the predicted efficiency of mixing processes should be treated with caution. However, it is a matter of fact that chemical anomalies due to pollution of extra-stellar material are actually observed in stars (see next Section) in agreement with recent studies predicting a long-lasting residue over-metallicity on the surface of $\gtrsim$1~M$_{\odot}$ stars \citep[e.g.][]{Sevilla22,Behmard23}. %Asteroseismic studies of these chemically anomalous stars will help derive to which level they are overmetallic down to the center or only in their outer layers and give an important hint for a better understanding of the efficiency of thermohaline convection.

\section{Observations of chemical signatures of planet engulfment events}
\label{Sec:obs}

\begin{figure*}
    \centering
    \includegraphics[width=1.\textwidth]{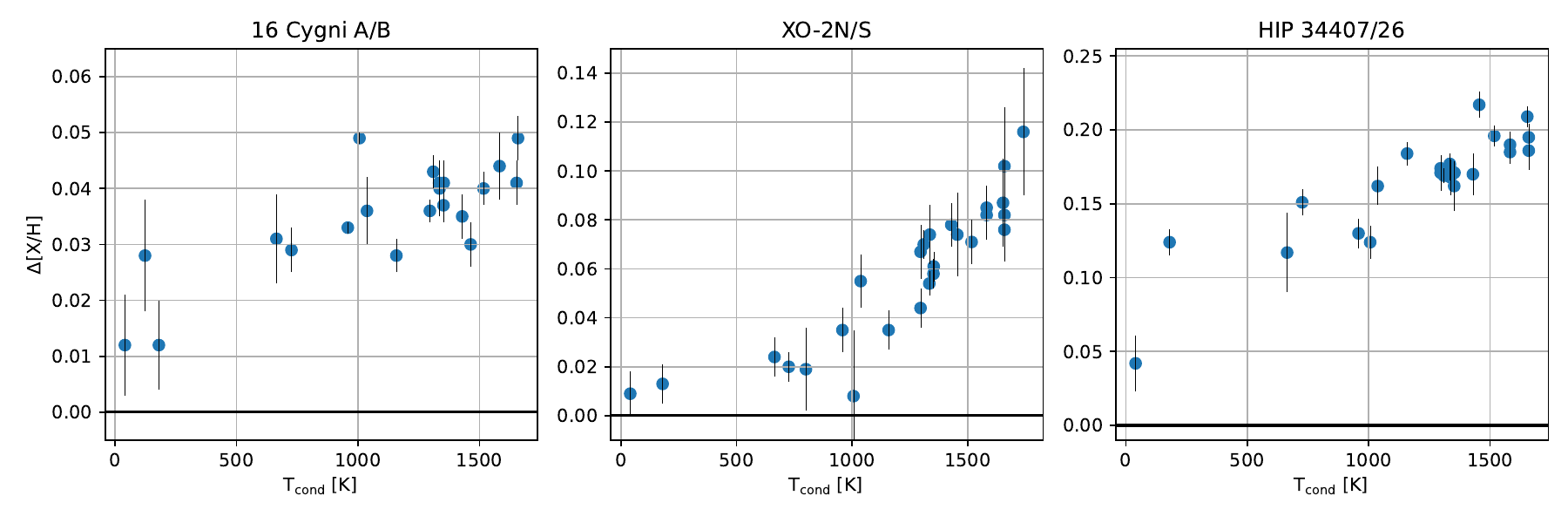}
    \caption{Differential abundances $\Delta$[X/H] between components of binary pairs as a function of the condensation temperature T$_{\rm cond}$. These trends are interpreted as chemical signatures of planet engulfment events. Data on 16~Cygni, XO-2N/S, and HIP~34407/26 are from \citet{Maia19},  \citet{Ramirez14}, and \citet{Ramirez19}, respectively.}
    \label{fig:anomalies}
\end{figure*}

\begin{figure}
    \centering
    \includegraphics[width=0.7\textwidth]{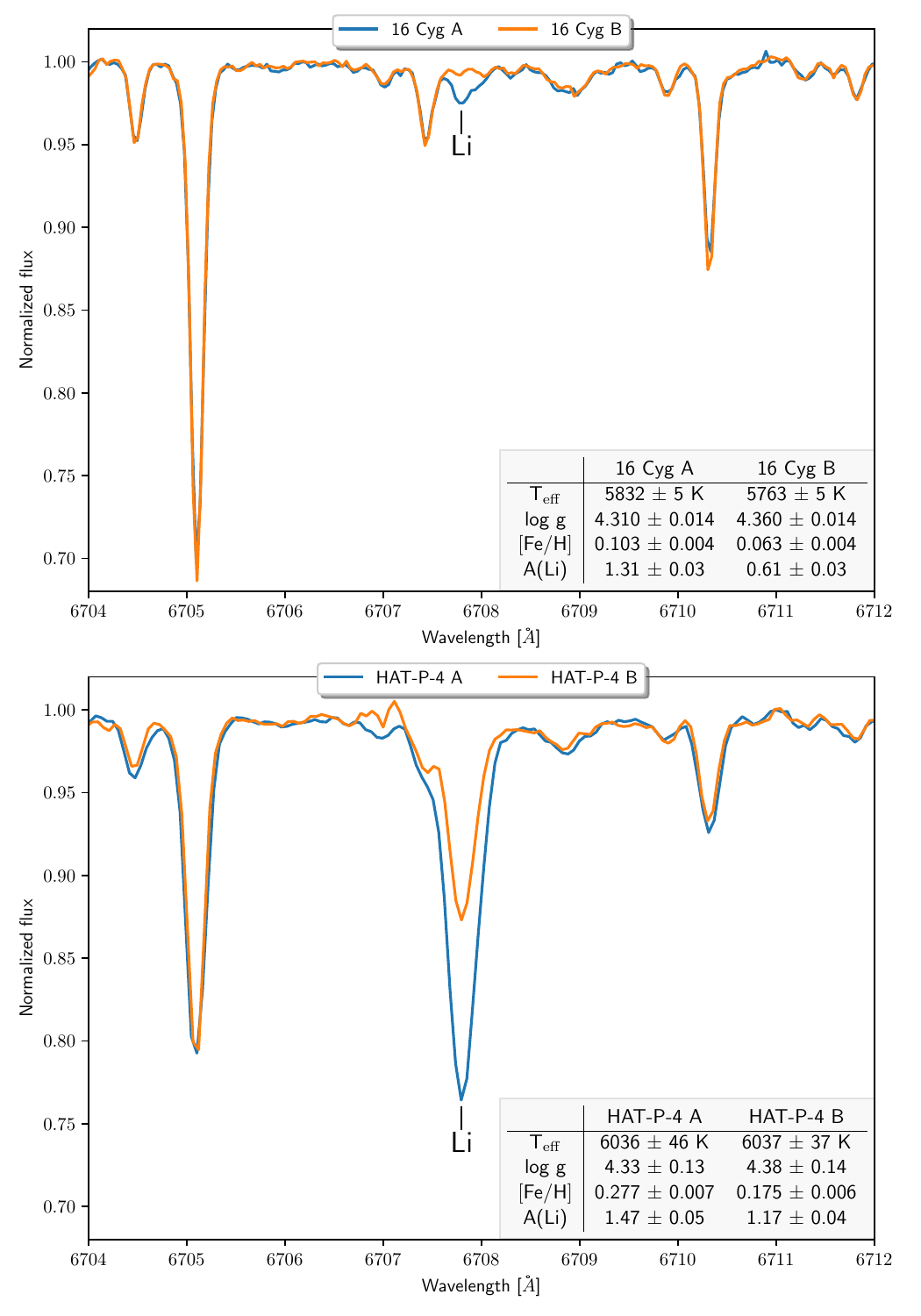}
    \caption{Normalized spectra of 16 Cyg A/B (top panel) and HAT-P-4 A/B (bottom panel) around the Li line at 6707.8~$\AA$. Both these binary pairs are significantly inhomogeneous in Li abundaces. The anomaly may be the consequence of a planet engulfment event. The spectra of 16 Cyg A/B shown in this plot have been acquired by HARPS-N, while their stellar parameters and Li abundances are from \citet{Maia19}. The spectra of HAT-P-4 A/B, the stellar parameters and Li abundances are from \citet{Saffe17}.}
    \label{fig:Li_line}
\end{figure}

The chemical pattern of a star is the result of a complex interplay of different processes. Therefore, it is extremely difficult to fully disentangle the chemical signature of planets from other major effects, such as the Galactic chemical evolution which is responsible of a great chemical diversity between stars formed at different times and locations within the Milky Way \citep[e.g.][]{Liu20,Biazzo22}. However, all that does not hold for stars in stellar associations, such as open clusters or binary systems. Their members have formed at the same time, from the same material, within the same molecular cloud, therefore they are expected to be chemically identical \citep{Ting15,Armillotta18}. Indeed, stellar associations have been proved to be the ideal targets to search for chemical signatures of planetary engulfment events in atmospheres of Sun-like stars \citep{Melendez14,Biazzo15,Ramirez15, Spina15,Saffe16,Teske16, Saffe17,Reggiani18,Spina18b,Maia19,Ramirez19,Nagar20,Galarza21,Liu21,Jofre21}. Typically, these signatures consist in larger levels of chemical inhomogeneity between two stars of the same association for those elements with higher T$_{\rm cond}$. Examples of these chemical anomalies observed among components of binary systems are shown in Fig.~\ref{fig:anomalies}. Some chemically anomalous pairs - but not all - also show significant differences in the Li abundances as well, with the metal-richer stellar component being also Li-richer. Two examples are shown in Fig.~\ref{fig:Li_line}. All these chemically anomalous pairs, along with many others showing similar features, are in agreement with what is expected from pollution of rocky bodies. In fact, these latter are particularly rich in refractory elements and Li (see Fig.~\ref{fig:rocky_composition}).
However, note that due to the depletion that normally takes place in young Sun-like stars, Li has been identified as a strong indicator of planet ingestion only if this latter occurs 30~Myr after star formation \citep[][]{Sandquist02}.

%not all pairs with a non-zero T$_{\rm cond}$-slope show variations in Li because pollution from non-stellar material is a necessary but non-sufficient condition for a Li over-abundance \citep[][]{Sandquist02}. Examples of chemical signatures of planet engulfment events observed among components of binary systems are shown in Fig.~\ref{fig:anomalies}. %The differential abundances $\Delta$[X/H] between the two stars of a pair are progressively more dissimilar as we move towards higher T$_{\rm cond}$, meaning that the objects that have polluted the stellar atmospheres and thus are responsible for the observed abundance variations were rich in refractory elements and poor in volatiles. In other words, these objects were probably formed by rocks.

For practical reasons, the level of chemical inhomogeneity between members of a stellar association is typically quantified by the slope between $\Delta$[X/H] and T$_{\rm cond}$, the so-called T$_{\rm cond}$-slope. However, this approach has some limitations because there could be important deviations from linearity on the distribution of elements within the $\Delta$[X/H]-T$_{\rm cond}$ diagram. For instance, notice in Fig.~\ref{fig:anomalies} that the $\Delta$[X/H]-T$_{\rm cond}$ distribution observed for XO-2N/S has a break at T$_{\rm cond}$$\sim$1000~K which is not present in the other two pairs. In fact, the exact shape of the $\Delta$[X/H]-T$_{\rm cond}$ distribution is presumably driven by the detailed chemical pattern of the accreted rocks, which in turn depends on the temperature at which they formed.  Also, one might be tempted to use the $\Delta$[X/H] values to infer the mass and the type of material ingested by the star \citep{Galarza21,Jofre21}. However, that is a speculative exercise because the $\Delta$[X/H]-T$_{\rm cond}$ distribution is significantly affected by other unknown variables, such as the size of the convection zone at the time of the engulfment event and the efficiency of the thermohaline convection. 

Unfortunately, beyond the mere existence of an oddity in the chemical pattern of a star, very little information can be achieved by studying these anomalous pairs separately, case by case. For instance, considering a single anomalous pair, such as one of those shown in Fig.~\ref{fig:anomalies}, is not sufficient to irrefutably demonstrate that planet engulfment events exist. In fact, broadly speaking chemical variations among members of the same stellar association could be the result of a myriad of processes, including chemical inhomogeneity within protostellar gas clouds. Instead, it is through statistical studies on large samples of binary systems that we can gather deeper insights about the nature of the observed chemical anomalies. A recent study by \citet{Spina21} on 107 binary systems formed by Sun-like stars has provided three pieces of evidence in favour of the planet engulfment scenario. 

\begin{itemize}

\item The occurrence of chemically anomalous systems increases with the average temperature of the pairs. The same conclusion is reached by \citet{Yong23} on an independent sample of 91 binary pairs. This evidence cannot be accounted for by inhomogeneities within the protostellar cloud. Instead, it proves that stellar convective zones were polluted by external material, leading to modifications in the atmospheric chemical compositions.

\item The metal-rich components of the anomalous pairs have refractories-over-volatile abundance ratios (i.e., [Fe/C]) that are typically higher than those of co-eval stars. Conversely, the metal-poor components have standard [Fe/C] values. Thus, among the two members of a chemically anomalous pair, it is the metal-rich star that is anomalously richer than the companion. It is not the other way around. Hence, the metal-rich star was polluted by external refractory-rich material.

\item In many cases the ingested material was richer in Li than the unpolluted stellar atmosphere. Also this final clue indicts ingestion of rocky material for being responsible of the chemical anomaly.
\end{itemize}

Beyond these evidences, \citet{Spina21} also established that planet engulfment events occur in stars similar to our own Sun with a probability ranging between 20 and 35$\%$. Note that these values may be a lower-limit as some signatures of planet engulfment events might be quenched by mixing processes, which are not considered in the analysis. 

In conclusion, a significant fraction of planetary systems orbiting stars similar to the Sun have undergone an exceptionally dynamic history, culminating with the infall of planetary material into the host star. This is in contrast to the history of our Solar System, which has maintained its planets in nearly circular orbits.

\subsection{The chemical composition of the Sun}

%How does our own Sun fit in the context described above? 

The makeup of our Solar System differs from most exoplanetary systems found to date, even after accounting for observational biases \citep[e.g.][]{Winn15,Horner20}. The stable circular orbits of the planets in our own Solar System suggest that they formed on paths similar to their present-day orbits, with no significant changes. In contrast, the majority of exo-solar systems exhibit minimal resemblance to our well organised Solar System as they often host planets that could not form in-situ, such as close-in super-Earths and mini-Neptunes \citep[e.g.][]{Howard10}, and many other objects showing evidence for past significant planet-planet interactions \citep[e.g.][]{Juric08,Carrera19,Bowler20}. Consequently, the peculiarity of the Solar System in the context of extra-solar planets appears to be intimately linked to two key processes: orbital migration and dynamical instabilities. However, as it is discussed in the previous sections, orbital migration and dynamical instabilities can lead to planet engulfment events which - in turn - can be responsible of specific variations in the chemical pattern of stars.

Given these premises, it is tempting to search for a possible connection between the unusually ordered architecture of the Solar System to the chemical pattern of the Sun in comparison with those of other stars: is the chemical composition of the Solar atmosphere anomalous? If so, can this presumed anomaly be somehow liked to the unusual architecture of the Solar System? 
Finding answers to these questions is an extremely difficult challenge because - contrarily to stellar components of binary systems - there are no stars against which we can compare our Sun to determine whether it has an anomalous composition or not. The best strategy one can possibly embrace is to compare our Sun to the other stars in the Solar vicinity. However, note that process of stellar migration and Galactic chemical evolution - if neglected - could lead us to wrong conclusions.

\begin{figure}
    \centering
    \includegraphics[width=0.7\textwidth]{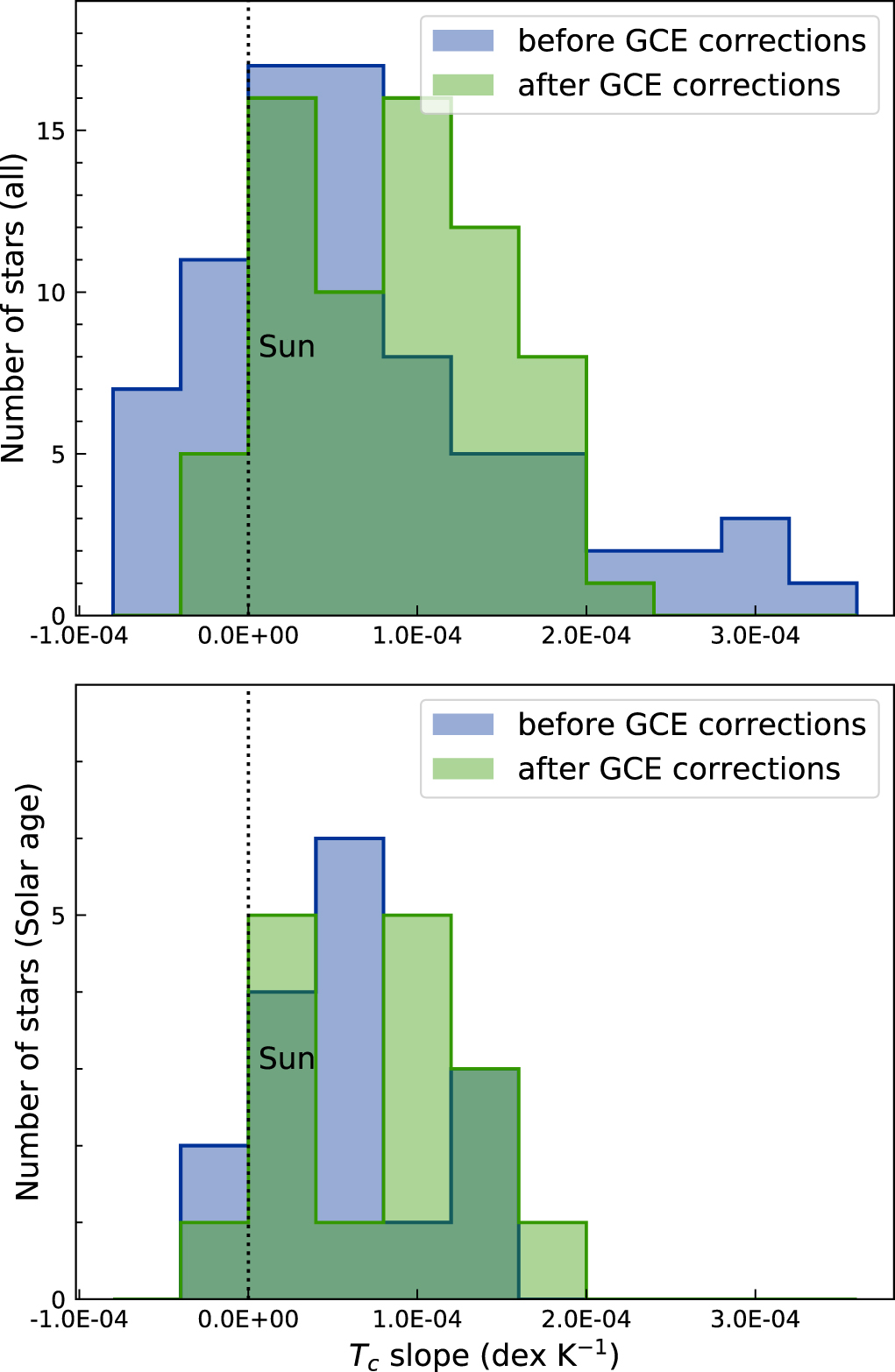}
    \caption{Distribution of T$_{\rm cond}$ slopes in the full sample of 79 Sun-like stars analyzed by \citet{Bedell18} (top) and in the subsample with ages between 3.5 and 5.5 Gyr (bottom). Slopes in [X/Fe] against T$_{\rm cond}$ are shown using abundances from before (blue) and after (green) correcting for the Galactic Chemical Evolution (GCE). The Solar T$_{\rm cond}$ slopes, which is zero by definition, lies below 93$\%$ of the sample after correcting for GCE. The Sun’s relative deficiency in refractory material may be related to the rarity of the Solar-System architecture among exoplanetary systems. The plots are from \citet{Bedell18}.}
    \label{fig:Tcond_Sun}
\end{figure}

Following this approach \citet{Melendez09} compared the chemical pattern of the Sun to that of 11 Sun-like stars finding that the Sun has indeed a $peculiar$ chemical composition, characterised by a lower refractory-to-volatile abundance ratio. The result is confirmed by \citet{Spina16b} and \citet{Bedell18} on larger samples of stars and accounting for effects of the Galactic chemical evolution (see Fig.~\ref{fig:Tcond_Sun}). A possible explanation of this result is that, contrarily to the Sun, most of the other stars host highly dynamic planetary systems and thus have chances of ingesting planetary material resulting richer in refractory elements than the Sun. In other words, our Sun appears to be poorer in refractory elements because it hosts a stable planetary system. However, note that it is still unknown the effect of radial migration on these studies.

\begin{figure}
    \centering
    \includegraphics[width=0.7\textwidth]{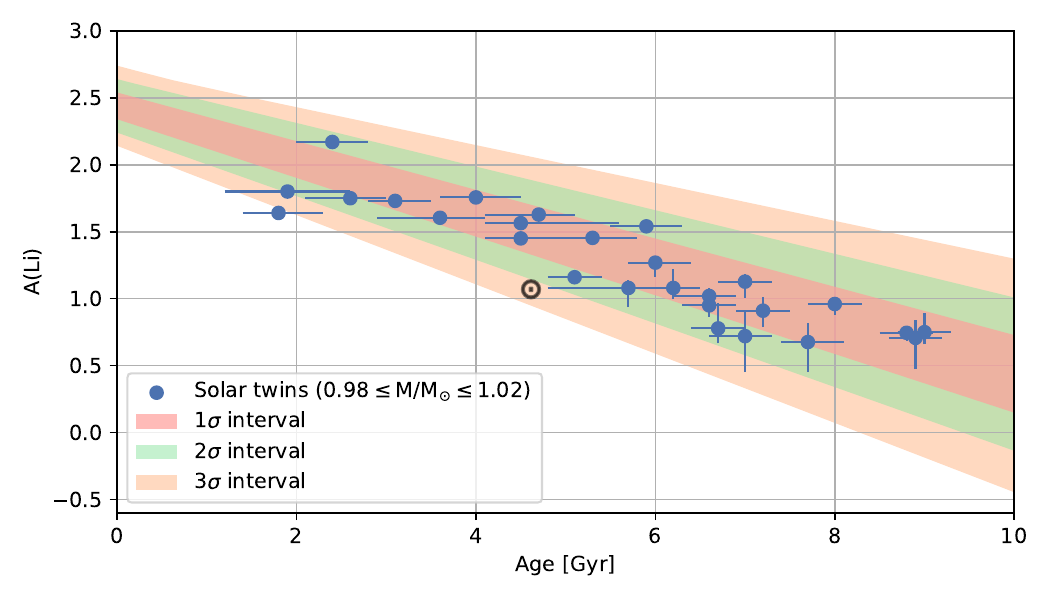}
    \caption{Lithium abundances versus stellar age as a function of mass for a sample of Sun-like stars with masses within 0.98 and 1.02 M$_{\odot}$. Confidence intervals of the linear distribution of stars are shown with coloured bands. The Sun’s relative deficiency in Li may be related to the rarity of the Solar-System architecture among exoplanetary systems. The plot is based on data published by \citet{Carlos19}.}
    \label{fig:Li_sun}
\end{figure}

Beyond the refractory-to-volatile abundance ratios, Li abundances bring additional clues within this framework. In fact, the amount of Li in the atmospheres of Sun-like stars is not affected either by Galactic chemical evolution, or by stellar migration. Instead, all stars with identical mass and age are supposed to have similar Li abundances. Only pollution from external material can generate a significant enhancement of Li in Sun-like stars. For that reason Li is considered a strong indicator of external pollution \citep[][]{Sandquist02}. A recent work by \citet{Carlos19} on a sample of stars with masses ranging between 0.98 and 1.02 M$_{\odot}$ has shown that the Solar Li abundance is lower more than two standard deviations that of co-eval stars (see Fig.~\ref{fig:Li_sun}).  This result as well is in agreement with a scenario where the Solar atmosphere has remained $unpolluted$ by the infall of rocks from the Solar System, while most of the other stars have undergone events of planet engulfment. 

\section{Alternative scenarios}

The previous sections of this manuscript describe how planet engulfment episodes can account for chemical anomalies seen among stars in associations. However, other scenarios have been proposed to explain these observations. In particular, it has been suggested that rocky planet formation itself may be able produce a deficit of refractory elements in planet hosting stars compared to others that have not formed rocky planets \citep[e.g.][]{Gonzalez97}. In fact, forming rocks implies a dust-gas separation within the protoplanetary disk: large amounts of refractory elements in the stellar surroundings are locked within the rocky bodies of the newborn planets while the volatiles are left free to accrete onto the star. On the other hand, a star that is not forming planets would indiscriminately receive both refractories and volatiles from the circumstellar disk. Therefore, planet forming stars are expected to become depleted of rocky forming material (i.e., refractories) than other stars without planets. However, this fascinating possibility has some pitfalls. First, assuming that the Sun is chemically anomalous because it has formed rocky planets implies that this process is rare around Sun-like stars, which is not supported by observations \citep[e.g.][]{Mulders18,Zink19}. Second, circumstellar disks dissipate on very short timescales: at ages of 5 Myr accretion onto the central star from the protoplanetary disk has already ceased \citep{Fedele10}. However, a 5 Myr old star of 1M$_{\odot}$ has a convective zone that still encloses more than 50$\%$ of the entire stellar material and, in order to produce a detectable Fe abundance variation of $\geq$0.05 dex, at least 100 M$_{\oplus}$ of rocky material are required (see Fig.~\ref{fig:enrichment}). This is certainly more than the mass in the rocky planets and asteroids of the Solar system. Third, the planet formation scenario cannot explain the anomalies in Li abundances that are observed in some binary pairs and in the Sun itself.

These first two pitfalls can be overcome by another scenario linked to planetary formation. Instead of considering the locking up of refractory material in rocky planets as the main mechanism, \citet{Booth20} proposed that refractories may be trapped in the outer protoplanetary disk by the cavities and gaps created by giant planets into the disk’s surface density. The existence of gaps and rings in protoplanetary disks is supported by observations \citep[e.g.][]{Andrews18}. When these gaps are located beyond the ice line of the refractory material, they can act as a barrier for these latter species. Namely, they can prevent refractory elements located beyond the gap in the outer disk and condensed into dust from falling towards the central star. Instead, the most volatile species which remain in the gaseous phase even in the outer disk can freely drift inward, cross the gap, and accrete onto the star. On the other hand, a protoplanetary disk without such cavities would enable the indiscriminate accretion of both refractory and volatile species. See Fig.~1 by \citet{Huhn23} for a very effective illustration of the mechanism. Thus, assuming that these cavities are due to the formation of giant planets, then stars hosting outer giant planets would have lower refractory-to-volatile abundance ratios than those of stars without outer giant planets. As it is shown by \citet{Booth20} and \citet{Huhn23}, the formation of outer giant planets can account for abundance variations of refractory elements up to $\sim$0.05 dex in the atmospheres of 1M$_{\odot}$ stars. Under this scenario, the peculiar refractory-to-volatile abundance ratio of the Sun would be linked to the architecture of the Solar System which hosts outer giant planets orbiting on nearly circular orbits. Whether this specific architecture is common or not around Sun-like stars is still matter of research. 

In conclusion, giant-planet formation is certainly a viable alternative to the planet engulfment scenario and it is possible that it has contributed to producing some of the observed anomalies in refractory elements. However, it should be noted that the variations of Li abundances detected in a number of binary pairs (e.g., see Fig.~\ref{fig:Li_line}) and the Sun’s relative deficiency in Li (see Fig.~\ref{fig:Li_sun}) are not predicted by the giant-planet formation scenario, instead they can be readily explained by planet engulfment events.

\section{Conclusions}
%The presence of chemical inhomogeneities between stars in associations is a well-established evidence \citep[e.g., ][]{Melendez14,Biazzo15,Ramirez15,Teske16, Saffe17,Reggiani18,Spina18b,Maia19,Nagar20}. In addition, there are observations suggesting that also the Sun may be chemically anomalous with a different refractory-to-volatile ratios when compared to other Sun-like stars \citep[e.g., ][]{Melendez09,Bedell18}. Although these latter could actually be the result of stellar migration, the possibility that our Sun is indeed anomalous is reinforced by the solar Li abundance which is lower than what is observed from co-eval stars \citep[][]{Carlos19}. All these anomalies show characteristic features that can only be explained by accretion of external material \citep{Spina21,Yong23} and are in excellent agreement with what is expected from planet engulfment events \citep[e.g.][]{Laughlin97,Pinsonneault01,Behmard23}. 

The presence of chemical inhomogeneities between stars in associations is a well-established evidence. These anomalies show characteristic features that can only be explained by accretion of refractory-rich material \citep{Spina21,Yong23} and are in excellent agreement with what is expected from planet engulfment events \citep[e.g.][]{Laughlin97,Pinsonneault01,Behmard23}. These observations carry substantial implications for the field of exoplanet science, as they are entirely detached from both the specific biases associated with exoplanet detection techniques and the assumptions made in n-body numerical simulations of planetary systems. First, a non-negligible portion of planetary systems orbiting stars similar to our Sun has undergone highly dynamic histories, ultimately resulting in the fall of planetary material into the host star. Second, the ability to identify chemical traces left by planetary engulfment events suggests that we can analyze a star's chemical makeup to determine whether its planetary system has experienced a highly dynamic history, in contrast to our Solar System. Consequently, we have a potential means to identify Sun-like stars that are more likely to harbour true analogues of the Solar System.

However, much work still needs to be done in this research field. First, observations of large controlled samples of binary systems would allow us to carry out statistical studies on the characteristic features of these chemical anomalies with comparison to models. That is necessary to estimate the frequency of these anomalies and establish their true nature. Second, it is necessary to assess whether or not the presence of these chemical signatures correlates with specific architectures of planetary systems. Finally, deeper investigations on the efficiency of thermohaline mixing should be carried out. In particular, sensitivity of current models to their initial assumptions should be tested. Also, asteroseismic studies of chemically anomalous stars will allow us to evaluate the level of enrichment down to the center.

%\section{Cross-References}

%\begin{itemize}
%\item Interiors and Surfaces of Terrestrial Planets and Major Satellites; Solar System–Exoplanet Synergies
%\item Disintegrating Rocky Exoplanets; Exoplanet Characterization
%\item Characterizing the Chemistry of Planetary Materials Around White Dwarf Stars; Exoplanet Characterization
%\item Characterizing Planet Host Stars: Introduction; Characterizing Planet Host Stars
%\item Planet and Star Interactions: Introduction; Planets and Their Stars: Interactions
%\item Signatures of Star-Planet Interactions; Planets and Their Stars: Interactions
%\item Rotation of Planet-Hosting Stars; Planets and Their Stars: Interactions
%\item Planetary Migration in Protoplanetary Disks; Formation and Evolution of Planets and Planetary Systems
%\item Formation of Giant Planets; Formation and Evolution of Planets and Planetary Systems
%\item Dynamical Evolution of Planetary Systems; Formation and Evolution of Planets and Planetary Systems
%\end{itemize}

\begin{acknowledgement}
It is a pleasure to acknowledge Anastasiia Plotnikova and Carlos Saffe for having shared with me the spectra of 16 Cyg A/B and HAT-P-4 A/B. 
\end{acknowledgement}

%  IF you do NOT use bibtex, put comments before the following 2 lines
\bibliographystyle{spbasicHBexo}  %for bibtex
\bibliography{HBexoTemplateBib} %for bibtex-example

\end{document}